\documentclass[a4paper,twoside]{article}
\newcommand{\affil}[1]{$^{\rm #1}$}
\usepackage[authoryear]{natbib}
\bibpunct{(}{)}{;}{a}{}{,}
\usepackage{graphicx}
\date{} 
\newcommand{\etal}{\mbox{\it{et al}~~}}
\title{\large\bf\flushleft The Bulge of M31}
\author{\parbox{\textwidth}{\flushleft
\vspace{-0.5cm}
{\it Jeremy Mould\affil{1,2}
\vspace{0.4cm}
{\small \affil{1}Centre for Astrophysics \& Supercomputing, Swinburne University,\ Hawthorn Vic 3122\\
 \affil{2}ARC Centre of Excellence for All-sky Astrophysics (CAASTRO)} 
\\
{\small Email: jmould@swin.edu.au}
}}}
\begin{document}

\begin{changemargin}{.8cm}{.5cm}
\begin{minipage}{.9\textwidth}
\vspace{-1cm}

\maketitle
%
%
\small{\bf Abstract: Bulges are not just elliptical subgalaxies situated in the centres
of large spirals. It might seem that way from their ages and chemistry, but bulge kinematics
have been known to be different since the first long slit spectra were obtained. M31 presents
the best opportunity to investigate all the issues of the stellar populations of bulges.
This review collects the array of probing data that has been accumulated in the last decade.
But the intriguing question `how did it form like this ?' remains.}

\medskip{\bf Keywords:} galaxies: bulges --- galaxies: abundances --- galaxies: formation --- galaxies: individual (M31)

\medskip
\medskip
\end{minipage}
\end{changemargin}

\small

\section{Introduction}

Data on our nearest neighbour bulge bearing galaxy have grown exponentially in the last few years and it is time they were reviewed. We begin by outlining the questions these data were assembled to address.
The first of these is how bulges form.  The classical answer is like elliptical galaxies. But since the ground-breaking review of Kormendy \& Kennicutt (2004) (KK2004) an alternate answer is, secularly from the disk. 
Whatever the reader's opinion on this, we could all agree that this is an interesting question, key to understanding the Hubble sequence.

This paper reviews the standard stellar populations clues to this question, the age distribution of bulge stars, i.e., the star formation history (SFH), the kinematics and the chemistry. The last review of bulges by Freeman (2008) noted various kinds of bulges, classical and boxy/peanut-shaped, pseudo bulges. He considered likely formation mechanisms, chemical evolution 
the relation between bulges and metal-poor halos (often lumped together as spheroids),
scaling laws for  bulges. One could add that an important new rationale for collecting and reviewing these data are as input to semi-analytic (SAL) models.
As the resolution of cold dark matter models improves, better and better semi analytic models will ride on them interpreting the baryon mass fraction. We need these SAL models to make realistic bulges and disks. 

\section{Bulge or Pseudo-bulge ?}
Is the M31 bulge a classical bulge, a pseudo-bulge or something in between ? KK2004 conclude that M31 -- along with the other nearby galaxies M81, NGC 2841, NGC 3115, and NGC 4594 -- have classical bulges. 
However, a current summary would be more nuanced. Internal secular evolution involves the buildup of dense central components in disk galaxies 
that look no different from classical merger built bulges, but were made slowly out of disk gas
(Sellwood 2000;  Norman, Sellwood, \& Hasan 1996; Sellwood \& Carlberg 1984;  Olle \& Pfenniger 1998).
KK2004 call these pseudobulges. They have flatter shapes, high v/$\sigma$, small $\sigma$ relative to the Faber-Jackson $L~\propto~\sigma^4$ expectation, 
bars and exponential surface brightness profiles. In other words, a quantitativefit to surface photometry is required to distinguish classical from pseudobulge, 
rather than a morphological examination.

A majority of bulges appear rounder than their disks. These include the classical bulges in M31, M81, and the  Sombrero. The nucleus and bulge of M31 are dynamically independent 
	(Tremaine \& Ostriker 1982). KK2004 state that, if the ratio of bulge to total light exceeds  a half, the galaxy contains a classical bulge. 
But new data by Beaton \etal (2007) show a boxy-bulge in the outer bulge. And  Courteau \etal (2011) have updated the structural parameters. Nowadays we would say that
M31 has a classical bulge with pseudobulge trimmings.

By comparison, the Milky Way clearly has a boxy/peanut-shaped bulge, as seen in the Two Micron All Sky Survey star counts (Lopez-Corredoira \etal 2005) and the COBE/DIRBE near-infrared light distribution (Dwek \etal 1995). The long axis of this bulge/bar lies in the Galactic plane. According to Ness \etal (2012) the distribution of stars associated with the boxy/peanut structure shows complexity at higher Galactic latitudes and the disk is the likely origin.

\section{Influential Surveys of M31 and its bulge}
We note the work with the Mayall 4m and Hale telescopes, in particular Hodge (1982) (The Atlas of Andromeda) and Mould \& Kristian (1986) \& 4SHOOTER (Mould 1986). The second of these papers introduced the tip of the red giant branch
as a distance indicator, and the third found 47 Tuc-like metallicity at the
bulge/halo interface.
The discovery of halo streams came with the survey with the Isaac Newton Telescope (Ibata \etal 2001), and now 
we have full stellar population photometric studies with PANDAS (McConnachie \etal 2009) and the Hubble Space Telescope's PHAT (Dalcanton \etal 2012 and Williams \etal 2012). 
The Keck Telescope DEIMOS multiobject spectrograph has produced the SPLASH Survey
of the Giant Southern Stream (Gilbert \etal 2009)  and (of relevance to the bulge) Dorman \etal's (2012) kinematics of the inner spheroid. Dorman \etal show
that velocity anisotropy contributes to the flattening of the spheroid
and that the inner spheroid/outer bulge resembles an elliptical galaxy in this respect.

At the same time we have seen multiwavelength surveys (Figure 1\footnote{http://dirty.as.arizona.edu/$\sim$kgordon/m31\_press/m31\_press.html}). Most recently, Groves \etal (2012) present Herschel telescope data. 
The dust emission tells us the distribution of the bulge's stellar population which is responsible for heating the dust.
Other surveys include an LPV survey (Mould \etal 2001), a Spitzer survey (Mould \etal 2008) and a Pan-STARRS survey (Chambers \etal 2009).

These surveys have influenced our understanding of M31's bulge, starting with the 4SHOOTER CMDs, which showed surprisingly high metallicities
at large radii, followed by the INT survey, which found the first evidence for a major merger, the Great Southern Stream and PANDAS, which saw
coherent structures that are almost certainly remnants of dwarf galaxies destroyed by the tidal field on M31. 
Dorman's, Groves's and Williams's work is most relevant to the bulge. The kinematic survey is discussed in the next section.

\subsection{The Stellar Population of M31's Bulge}
Three important questions are 
\begin{itemize}
\item What is its SFH, continuous or initial burst, and how much of each ?
\item What is its chemical enrichment history; is it similar to the Milky Way bulge ?
\item What is its relation to the other components, the halo, the globular cluster system, and the disk ?
\end{itemize}

A powerful answer to the first question is provided by HST (Brown \etal 2009 and references therein). In a field 11 kpc on the SE minor axis Brown \etal (2003) find that
the M31 halo contains a major ($\sim$30\% by mass) intermediate-age (6--8 Gyr) metal-rich ([Fe/H] $>$ --0.5) population, as well as a significant globular cluster age (11--13.5 Gyr) metal-poor population. Ferguson \etal (2005) have pointed out that this field is rather close to significant substructure/tidal debris and the derived SFH may be special.
To the extent that, at many r$_e$, this samples the bulge, we might say that M31 hosts a traditional bulge.
Brown \etal (2004) show a CMD for a bulge globular cluster and find an age in the range 9--12.5 Gyr.
 Deep photometry of the M31 halo shows populations of ancient metal-poor stars, 
one extending to younger ages and high metallicity, apparently due to its active merger history.

The subdominant intermediate age population is highlighted by the AGB luminosity function.
By the fuel burning theorem (Greggio \& Renzini 2011) a flat/declining AGB luminosity function corresponds to a flat/declining star formation rate
over timescales of Gyrs, the precise starting and endpoints of which are given empirically by Frogel, Mould, \& Blanco (1990) for a Magellanic Clouds composition,
or theoretically by the Padua isochrones (Bertelli \etal 2008). The calibration is mass loss rate dependent
(Marigo \etal 2010; Melbourne \etal 2010; Girardi \etal 2010).
Similarity of the AGB luminosity function (LF) in M31's inner bulge and Baade's window is seen both in Davidge's (2001) work and NICMOS (Stephens \etal 2003).
This implies that both bulges are mostly as old as globular clusters with an admixture of younger stars (e.g. Clarkson \etal 2011, Rich 1999).

The M31 bulge has a central metal rich component that can be seen in the JHK CMD of PHAT. If we compare 
B1 = central bulge and B9 = outer bulge, we see in Figure 2 very red stars at the tip of the RGB. These are metal rich. 
Differences in the giant branch morphology at the faint end are due to confusion in Brick 1. 
Other evidence for the similarity of the metallicity distributions of M31's bulge and the stellar population of Baade's window is the metallicity distribution of the planetary nebulae.

Lauer \etal (2012) find that the black hole in M31's dual nucleus is surrounded by a flattened cluster of blue stars.
Similar blue stars are seen over the central pc. These may be young stars or they may be post-HB stars.
The Milky Way also has central blue stars (Rafelski \etal 2007 )for which youth has been demonstrated, but it also has a central molecular zone (Shetty \etal 2012), which M31 lacks.
An Occam's razor approach would suggest that, while the M31 star cluster may indeed have massive young stars, this is not proven for the central bulge, whose stellar population may have nothing much younger than a billion years, as suggested by its AGB LF.  

We turn to the kinematics. The classical work is the thesis of McElroy (1983). More recently Saglia \etal (2010) presented
new optical long-slit data along 6 position angles of the bulge region of M31. 
They found the velocity dispersions of McElroy (1983) to be underestimated and that previous dynamical models underestimated the stellar mass of M31's bulge by a factor 2. 
They noted that the velocity dispersion of M31 grows to 166 km/s, thus reducing the discrepancy in the mass of the BH at the centre of M31 with the Magorrian relation.
They showed that the kinematic position angle varies with distance, pointing to triaxiality and found gas counterrotation near the bulge minor axis. From
Lick indices of long slit spectra at several position angles in the inner
few arcmin Saglia \etal found
the bulge of M31  (except for the inner arcsecs of the galaxy) to be uniformly old ($>$12 Gyr) at solar metallicity with [$\alpha$/Fe] $\approx$ 0.2 and 
approximately radially constant M/L$_R~\approx$  4, in agreement with stellar dynamical estimates. The AGB LF measures discussed earlier would be more sensitive
to a sprinkling of billion year old stars that these integrated light samples.
Saglia \etal found that
in the inner arcsecs the luminosity-weighted age drops to 4--8 Gyr, while the metallicity increases to above 3 times the solar value.

\subsection{X-ray emission}
Gilfanov and Bogdan (2010) found that
unresolved X-ray emission from the bulge of M31 is composed of at least 3 different components: 
\begin{itemize}
\item Broad-band emission from many faint sources, mainly accreting white dwarfs and active binaries, 
similar to Galactic Ridge X-ray emission of Milky Way. 
\item Soft emission from ionized gas with temperature $\sim$ 300 eV and mass $\sim$ 4 $\times$ 10$^6$ M$_\odot$. 
The gas distribution is significantly elongated along the minor axis, suggesting that it may be outflowing perpendicular to the disk. 
\item Hard unresolved emission from spiral arms, most likely associated with protostars and young stellar objects located in the star-forming regions.
\end{itemize}

It is the first of these that is the signature of the bulge stellar population, namely old and predominantly low mass stars.
With an effective bulge potential, GM/r$_e$ of 10$^5$ (km/sec)$^2$ (10$^7$K), supernovae heating would readily drive
a wind of bulge gas, evidenced by the second component.
The ram pressure of the wind would find lower resistance towards the galactic poles, but the elongation of the low energy xrays is modest\footnote{http://chandra.harvard.edu/phot/2006/m31}.

\subsection{Formation theory}
Binney (2007) sums up formation theory with the twin ideas that pseudobulges form from unstable disks, while classical bulges form in violent episodes of star formation
when a merger sweeps cold gas into a galactic centre. For classical bulges Steinmetz (2003) suggested rapid infall in cold dark matter models at redshift around 6.
Bekki (2010) specifically simulated the globular cluster (GC) system of M31 and
found that a major merger event could have formed bulge and rotational kinematics of the system.
He numerically investigated the kinematics in both disk and halo components.
It shows maximum rotational velocities of 140--170 km/s, 
and for a range of orbital parameters of merging
there is a rotating stellar bar (a boxy bulge) which can be formed in models for which the GCs show strongly rotational kinematics. 

\subsection{Chemistry}
Jablonka \& Sarajedini (2005) show the 
giant branch chemistry of the bulge.
With individual K giant stars affected by confusion, this may be the best probe of metallicity available.
Mould \& Spitler (2010) show (in an application to the Sombrero galaxy) that a metallicity distribution function can be analyzed to yield the SFH and accretion history,
using Z(t) as a proxy for time. The assumptions are a single zone (normally, `closed box' or `simple') model 
with infall of low metallicity gas.
This is shown for the M31 bulge in Figure 3. A period of rapid accretion early on
is the requirement for this fit. Note that the Mould \& Spitzer model is $not$ a closed box model; it is a leaky
box model.

\subsection{Scaling relations}
The scaling relations for bulges are intensely interesting. Virial equilibrium places galaxies in a 3-space of mass, radius and a dynamical parameter.
Mass and effective radius are well defined, but what is the relevant dynamical parameter for bulges?
Catinella \etal (2012) suggest that that is a combination of central velocity dispersion and disk rotation. 
Pizzella \etal (2005) show relations between central velocity dispersion and disk rotation. 
M31 fits well in the former (GASS) sample of massive galaxies. 
SAL models suggest that the presence of a large bulge, 
like M31's, affects position in the Tully-Fisher relation (Tonini \etal 2013).
Tonini \etal examine $$L \propto (\Delta V(0)^4 + c \sigma^4)^{1/4}$$ where c is a constant.
More data are being obtained in a follow up of the HIPASS sample of Sa and Sb galaxies by 
Mould, Webster, Freeman, Tonini, Jones and collaborators.

\subsection{Bulge mass and light}
 Widrow, Perrett \& Suyu (2003) found that
the most successful models in a $\chi^2$ fit assume a bulge mass that is nearly a factor of 2 smaller than the value from Kent (1989) of 4 $\times$ 10$^{10}M_\odot$. 
Their galaxy model with bulge mass 2.5 $\times$ 10$^{10}~ M_\odot$  and disk mass  7 $\times$ 10$^{10}~ M_\odot$  provides a good overall fit to observations, and yields 
M/L ratios that are quite acceptable and stable against bar formation (see also Widrow \& Dubinski (2005).

Irwin \etal (2005) measured effective V- and i-band minor-axis profiles. The V-band profile is illustrated in green and blue, and the i-band profile in black and red. The data were derived from surface photometry and star counts. 
A de Vaucouleurs R$^{1/4}$ law was fitted with b$_{eff}$ = 1.4 kpc. The dashed black line in the left-hand panel 
of Figure 4 shows an NFW profile computed with a scale radius of 3.4 kpc and, in the right-hand panel, 
an exponential profile was computed with a scale length of 13.7 kpc to model a very slowly varying stellar halo.

Courteau \etal (2011) provide a basic photometric model for M31 which has a
 Sersic bulge with shape index n = 2.2 $\pm$ 0.3 and effective radius R$_e$ = 1.0 $\pm$ 0.2 kpc, 
and a dust-free exponential disk, scale length R$_d$ = 5.3 $\pm$ 0.5 kpc; 
parameter errors reflect a range of decompositions.
The bulge parameters are rather insensitive to bandpass effects. Courteau remarks that the 
value suggests a first rapid formation via mergers followed by secular growth from the disk.

\subsection{A bulge transient}
Mould \etal (1990) discovered
M31 RV, a luminous red variable star, somewhat brighter than the AGB tip, that appeared in the bulge of M31 in 1988. 
During outburst, which lasted a few months, it was one of the brightest stars in the Local Group. 
Unlike a classical nova, it was extremely cool during the eruption, and it never became optically thin or exposed a hot, blue source. In retrospect, there is a
remarkable similarity to V838 Mon, a luminous Galactic variable star that underwent a similar rapid expansion to become a red supergiant (Bond 2011). The
outburst mechanism for this new class of luminous transients remains unknown.

According to Bond \etal (2012) intermediate-luminosity red transients (ILRTs) are a recently recognized class of stellar eruptions with 
maximum luminosities between those of classical novae and supernovae. During their outbursts, which generally 
last a few months, they typically evolve to extremely red colours, completely unlike novae.
Besides M31 RV, examples include V838 Mon, SN 2008S, V1309 Sco, the M99 optical transient of 
2010, and the 2008 and 2010 ILRTs in the nearby spiral NGC 300.  
At present, it appears that there are (at least) two separate evolutionary channels leading to ILRT outbursts: 
mergers of close binaries (accounting for ILRTs in old populations and possibly V838 Mon), and eruptions on stars 
of about 8-12 M$_\odot$, possibly due to electron-capture SNe (accounting for ILRTs in young populations). 

M31 RV was 3 arcmin from the nucleus on the SE minor axis and is most likely to have been a bulge event. 
According to Figure 4 the bulge SB at M31 RV is V $\sim$ 20. The object is $\sim$600 pc from the centre of the disk of M31 and on the minor axis.
The disk light is therefore 2 disk scale heights fainter than B = 21.6. The odds are more than 10:1 that it is a bulge object.

\section{Isolating the main sequence in M31}
Discussing integrated light spectroscopy of stellar populations, Mould (2012) asks,
why observe giant stars in galaxy halos closer than Virgo, when they are nothing but trouble ?
There are horizontal branch (HB) anomalies, convective envelopes, mass loss etc.
This is the rationale for studying main sequence integrated light, rather than an integral over $all$ luminosities.
According to the fuel burning theorem,
the number of stars in any evolutionary stage is proportional to the luminosity of that stage
multiplied by its lifetime (N = bLt (Renzini \& Buzzoni 1984), where t is the lifetime of an evolutionary phase.) 
If you observe 200 solar luminosities in an old stellar population with a spectrograph or imager,\\
N(HB) = 2 $\times$ 10$^{-11} ~\times$  200 $\times$ 10$^8$ = 0.4 stars, and\\
N(MS) = 0.4 $\times$ 1.3 $\times$ 10$^{10}/10^8$ = 50 stars.\\
There are also RGB stars and subgiants to mask out along with the HB stars, but they are weak in the blue.

\subsection{The potential of disintegrated spectra}
We may refer to such IFU spectroscopy as disintegrated or disaggregated to distinguish it from the analysis of integrated light.
Rapid collapse models of galaxy formation predict small gradients in mean age and large gradients in metallicity.
In dwarf merger models one expects large age variance and an accompanying metallicity range.
Separate measures of mean age and mean metallicity of simple stellar populations can
distinguish these models and constrain parameterized examples of such models. 

But there are other potential applications. The low mass end of the
initial mass function from FeH Wing-Ford band may be measureable without contamination from M giants.
The Ba/Fe or Ba/Ca ratio would indicate whether the $s$-process had been active in the extreme halo.
Ca triplet and Paschen lines behave similarly and are addressable with large telescope adaptive optics.

\subsection{M31: start with the halo, move on to the bulge}
The potential outlined above motivates making full main sequence evolutionary synthesis models.
Effectively this means integrating Mould's (2012) results right down the main sequence.
A trial run on the M31 halo with an IFU spectrograph can be outlined as follows for
HB stars at B = 25 mag.\\
Suppose the surface brightness is $\mu_B$ = 24 mag/arcsec$^2$, $cf$ sky = 22 mag/arcsec$^2$.\\
In 1 arcsec$^2$ log L/L$_\odot$ = 0.4(24.35--24 ~+~ 5.4) = 2.3, i.e. the 200 L$_\odot$ of $\S$4. \\
In 100 arcsec$^2$ there are 100 $\times$ 0.4 = 40 HB stars to be masked.

In the M31 bulge this technique would only be useful in the outer lower SB areas with the milliarcsec
resolution of extremely large telescopes.
Of course, no spectrum need be wasted. Sorting the fibres by luminosity when the datacube is produced
(schematically shown in Figure 5) will allow one to study HB light and MS light separately and everything
in between.


\section{Summary}
The M31 bulge is the closest bigger bulge galaxy than the Milky Way. 
Understanding its formation presents two worthy challenges:
collecting the right data, and using them to constrain detailed formation models.
What we know at present is that 
its stellar population is mostly old stars. Some are middle aged and these tend to be centrally concentrated.
The structural parameters of the bulge are those of a classical bulge in the central area,
but there is evidence for some sort of bulge-disk crosstalk in the outer parts, suggesting secular evolution of some kind.
 
The SFH should be measured in the centre and at a radius where a kinematic filter can isolate bulge stars.
The bulge has a broad metallicity distribution, but the data are rather limited compared with what could be achieved spectroscopically.
This metallicity distribution yields the gas accretion history and the SFH.
Maps of mean age and metallicity are realizable from the turnoff and giant branch respectively,
either by star counts or disintegrated light. 

The question as to 
the primary influence on formation remains (Obreja \etal 2012). Was it the disk, the halo, or mergers ? 
A program to answer this question would involve at the very least,
\begin{itemize}
\item simulating secular disk processes to make predictions for the stellar population
of such a component of bulges: their v/$\sigma$, metallicity and age distributions,
\item simulating baryon dominated hydrodynamic collapse to make predictions for possible stellar population gradients, and
\item simulating mergers in gas and dark matter to compare with IFU maps, like that of SAURON\footnote{http://www.strw.leidenuniv.nl/sauron/}, of the M31 bulge.
\end{itemize}

\section*{Acknowledgments} This paper is based on a presentation to Martin Schwarzschild Centenary meeting at Princeton in June 2012. 
I would like to thank the organizers for their invitation to give this review. I appreciate the help of anonymous referees
in transitioning from the verbal to the written medium. The study of the evolving universe
is supported by the Australian Research Council through CAASTRO\footnote{www.caastro.org}.
The Centre for All-sky Astrophysics is an Australian Research Council Centre of Excellence, funded by grant CE11E0090.

\pagebreak

\begin{figure}[h]

\begin{minipage}[b]{0.6\linewidth}
\centering
\includegraphics[scale=2.5, angle=0]{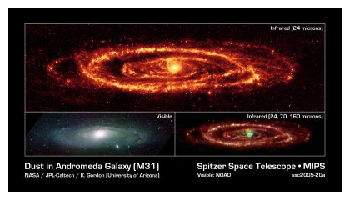}
\caption{Left: Spitzer MIPS images of M31.{\it NASA press release.} 
Right: Herschel-XMM images of M31 {\it ESA press release.} }

\end{minipage}
\hspace{.5cm}
\begin{minipage}[b]{0.4\linewidth}
\centering
\includegraphics[scale=1., angle=0]{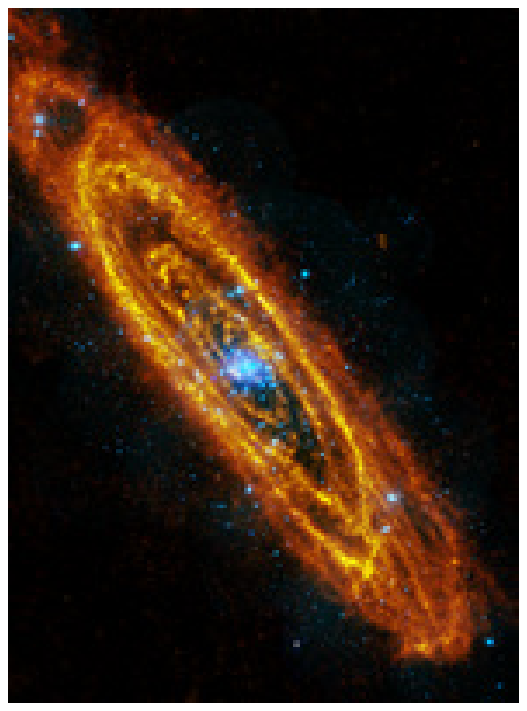}
\end{minipage}
\end{figure} 

\begin{figure}[h]
\vspace{1 truein}
\begin{center}
\includegraphics[scale=1., angle=0]{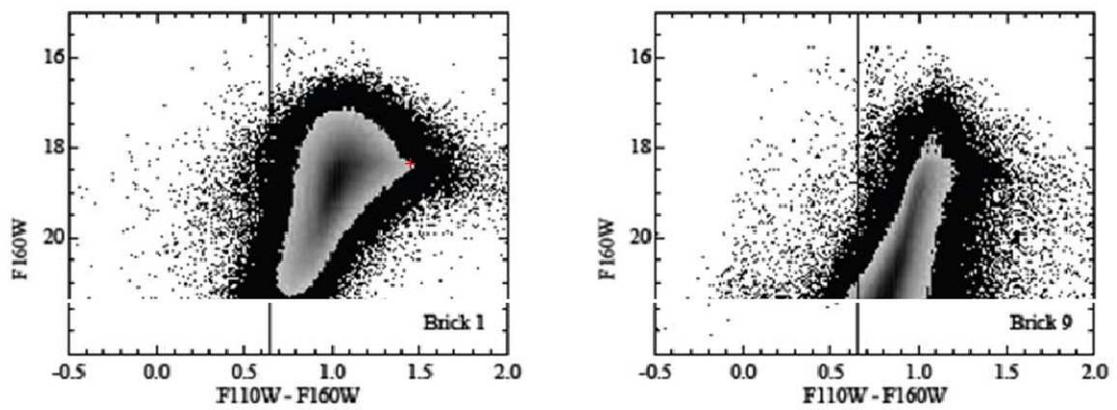}
\caption{The CMDs of brick1 and brick 9 in PHAT. In the inner bulge (brick 1)
very red and metal rich stars are seen at F160W = 18.5 mag. The metal rich feature 
is shown with a red plus.}
\end{center}
\end{figure}

\begin{figure}[h]
\begin{center}
\includegraphics[scale=.7, angle=-90]{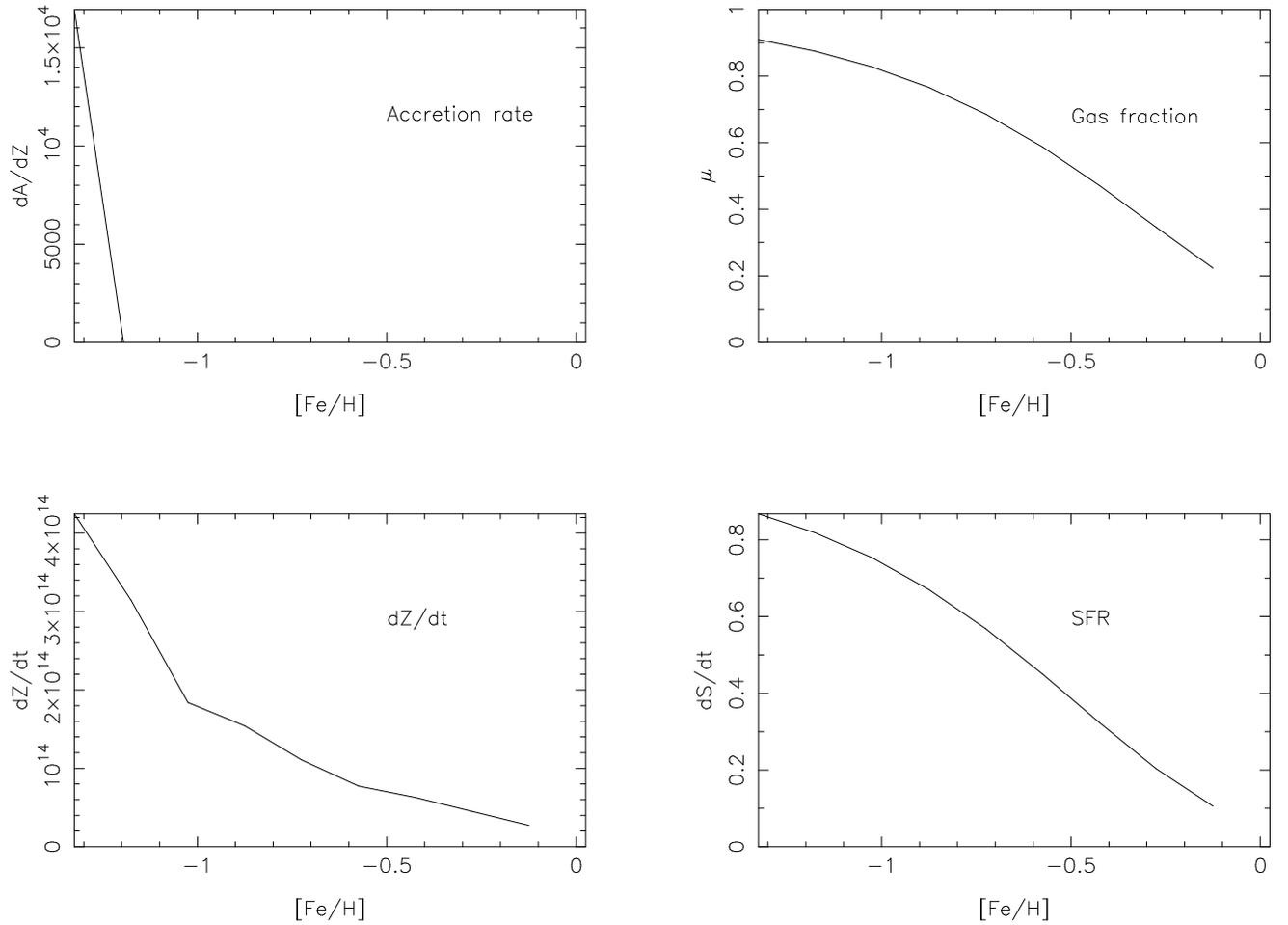}
\caption{The chemical enrichment model of Mould \& Spitler (2010) is a fit to the Jablonka \& Sarajedini (2005)
metallicity distribution of the M31 bulge 
In this model Z is a proxy for time, and so these plots
show accretion rate and star formation history as a function of time.}
\end{center}
\end{figure}

\begin{figure}[h]
\begin{center}
\includegraphics[scale=1., angle=0]{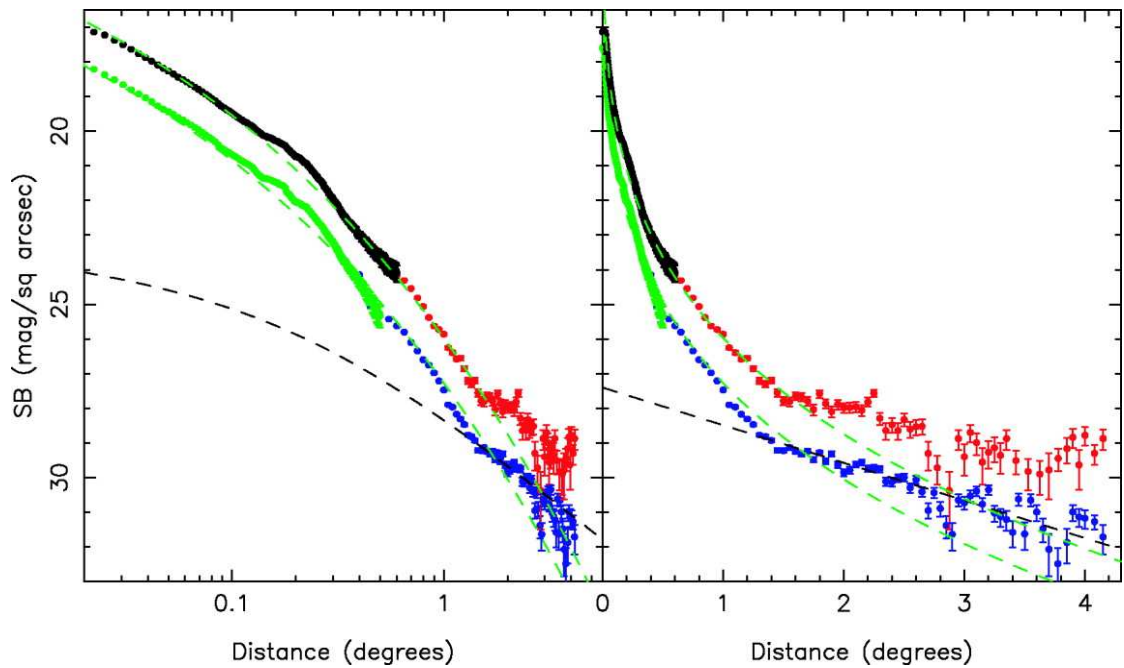}
\caption{Effective V- and i-band minor-axis profiles shown on a log-log (left) and log-linear 
(right) scale. The V-band profile is illustrated in green and blue, and the i-band profile in black
 and red. The green and black circles are derived from surface photometry, whereas the 
blue and red points are derived from star counts. The error bars reflect a combination of 
Poissonian and background uncertainties. The green dashed lines show a de Vaucouleurs 
R$^{1/4}$ law. The dashed black line in the left-hand panel shows an NFW profile computed 
with a scale radius of 3.4 kpc and, in the right-hand panel, an exponential profile computed 
with a scale length of 13.7 kpc.
Irwin \etal 2005, reproduced by kind permission of the author.}
\end{center}
\end{figure}

\begin{figure}[h]
\begin{center}
\includegraphics[trim=10mm 0mm -10mm 0mm, scale=.7, angle=-0]{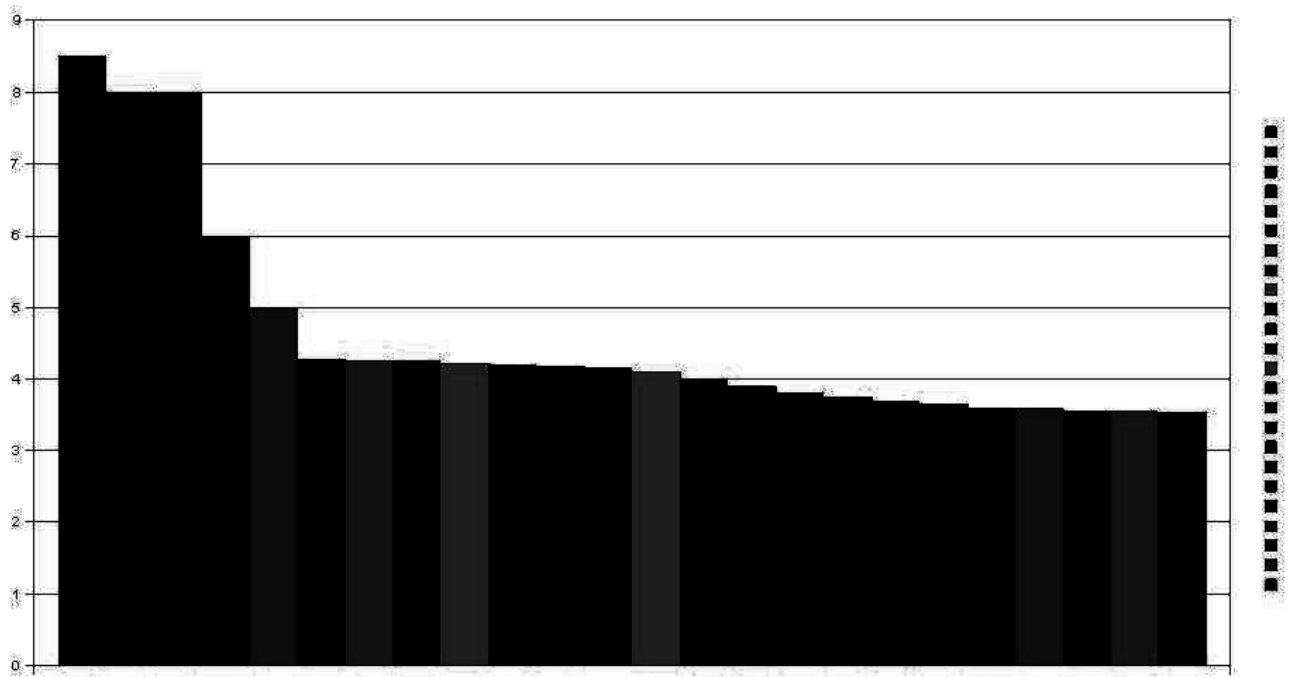}
\caption{In this imaginary datacube from an IFU spectrum of the M31 halo one HB star has illuminated the 4 fibres on the left.
The fibres to the right (not shown at their true frequency or flux) are MS light.}
\end{center}
\end{figure}


\end{document}